\documentclass[reprint,
 amsmath,amssymb,
prb,
floatfix,
]{revtex4-2}

\usepackage{graphicx}
\usepackage{dcolumn}
\usepackage{bm}
\usepackage{physics}
\usepackage{hyperref}

\begin{document}

\preprint{APS/123-QED}

\title{Enhancement of spin-to-charge conversion of diamond NV centers at ambient conditions using surface electrodes}

\author{Liam Hanlon}%
\affiliation{%
Department of Quantum Science and Technology, Research School of Physics, Australian National University, Canberra, ACT, Australia, 2601
}%
\author{Michael Olney-Fraser}%
\affiliation{%
Department of Quantum Science and Technology, Research School of Physics, Australian National University, Canberra, ACT, Australia, 2601
}%
\author{Lukas Razinkovas}%
\affiliation{%
Center for Physical Sciences and Technology (FTMC), Vilnius LT-10257, Lithuania
}%
\author{Marcus W. Doherty}%
 \email{mwdoherty@anu.edu.au}
\affiliation{%
Department of Quantum Science and Technology, Research School of Physics, Australian National University, Canberra, ACT, Australia, 2601
}%

\date{\today}

\begin{abstract}
The nitrogen-vacancy (NV) center in diamond is a heavily studied defect due to its potential applications to quantum metrology and computation, particularly in ambient conditions. The key mechanism to using the NV in any application lies in the ability to read out the spin state of the defect which is typically done optically. The optical contrast is then the key metric for electron spin readout fidelity and one of the key limiting factors in the NV's overall performance. We present a new mechanism for high contrast readout using the spin-to-charge conversion (SCC) mechanism in conjunction with an electrode to improve the spin contrast by altering the NV energy levels relative to the diamond conduction band. Theoretical modelling predicts an optical spin contrast at 42\% which would be the highest optical contrast for the NV at room temperature and the technique opens up a range of alternative research pathways for the NV which are discussed. 

\end{abstract}

\maketitle

\section{Introduction}\label{intro}

The nitrogen-vacancy (NV) center defect in diamond shows great promise as a tool for a variety of quantum applications including metrology and imaging \cite{Barry2016b, Kucsko2013, LeSage2013OpticalCells, Barson2021NanoscaleSpin, Lesik2019MagneticCenters} as well as computing and networking \cite{Maurer2012Room-temperatureSecond, Wu2019AConditions, Nizovtsev2005ASpins, Childress2013DiamondNetworks, Pezzagna2021QuantumDiamond}. The defect is easy to engineer, can be manipulated with simple laser and microwave pulse sequences and can be designed in a variety of forms by shaping the diamond structure that houses the defect \cite{Doherty2013TheDiamond, Babinec2010a, Siyushev2010, Appel2016a, Wan2018}. Arguably the most impressive quality of the NV center is its ability to initialise, manipulate and readout quantum states in ambient conditions \cite{Doherty2013TheDiamond}. This capability vastly increases the NV's applicability compared to other technologies, allowing for (among many other applications) powerful sensors of biological samples and room temperature quantum computers.  

One of the major drawbacks of the NV center is its low electron spin optical contrast and associated readout fidelity. The low contrast limits the sensitivity of an NV sensor and the overall computational fidelity of an NV quantum computer. In addition to this, under optical illumination, the NV can photoionize, changing its charge state from NV$^-$ to NV$^0$, a state that doesn't have the same optical spin initialisation and readout mechanisms. In this case, the NV needs to be converted back into the NV$^-$ state via optical recombination from the diamond valence band \cite{Aslam2013Photo-inducedDetection}. However, this ability to alter the NV charge state allows for the development of the spin-to-charge conversion (SCC) protocol, where the NV spin state is mapped to the NV charge state and the spin state is read out via a much higher fidelity charge state readout protocol. There have been many SCC protocols developed at cryogenic temperatures \cite{Hanlon2021Spin-to-ChargeDiamond, Irber2021RobustDiamond, Zhang2021High-fidelityConversion} and in ambient conditions \cite{Jaskula2019ImprovedConversion, Hopper2016Near-infrared-assistedDiamond}. Cryogenic SCC protocols perform better, but cannot work in ambient conditions, removing one of the NV's most useful advantages. Whilst the ambient SCC protocols are more applicable, their improvement is modest and does not address the issue of unintended ionisation for true charge state control in the NV. 

\begin{figure*}[!ht]
            \centering
            \includegraphics[width = 0.9\textwidth]{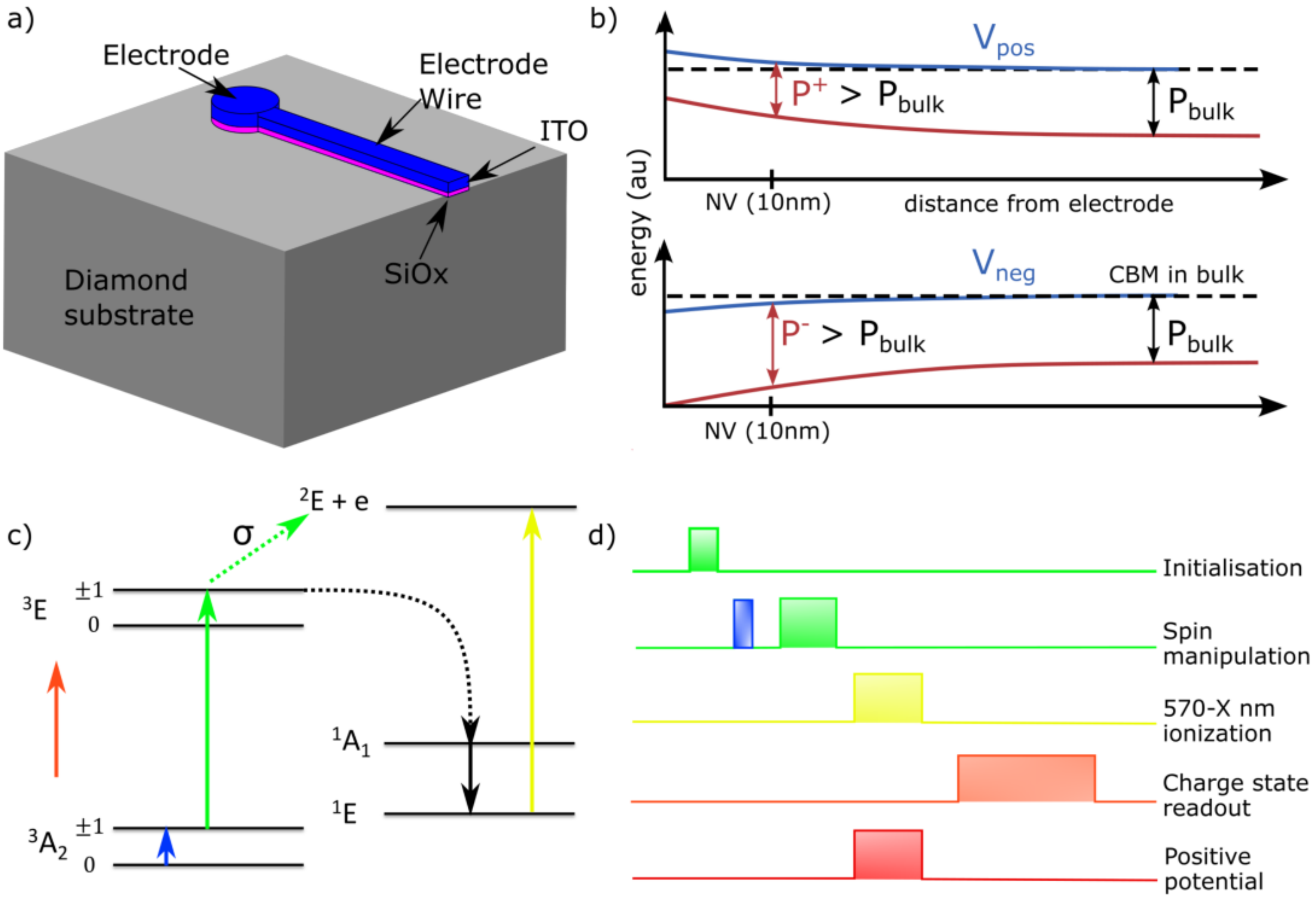}
            \caption{a) Image of the electrode over the diamond substrate where the main cylindrical electrode is over a near-surface NV (10~nm deep) and the electrode wire connects to a power supply which provides the electrode potential. b) Diagram of the electrode effect on the energy structure. The first image is with a positive electrode potential ($V_{pos}$) which shifts the conduction band minimum (CBM) and NV energy levels upwards compared to the CBM in bulk diamond with no potential, shrinking the photoionization energy gap $P^+$ compared to the bulk case with no potential ($P_{bulk}$). In the second image a negative potential shifts the energy levels downwards compared to the bulk CBM, widening the energy gap $P^-$. c) Energy diagram depicting the SCC protocol. The NV is initialised into the $^1$E singlet state with a green laser and microwaves (blue) where it is then ionised with a high power yellow laser into the NV$^0$ ${}^2$E state with an electron in the diamond conduction band. The charge state of the NV is then readout with a 595~nm orange laser pulse. This sequence also features the quadratic ionisation from the ${}^3$E state to the conduction band which has a cross section ratio of $\sigma$. The pulse sequence for the same SCC protocol in c) is shown in d) and includes the pulsing for the electrode itself, where the positive potential is turned on during the ionisation phase.}
            \label{fig:AmbientSCC}
        \end{figure*}

We introduce a new SCC protocol at ambient conditions with the application of an electrode over a near-surface NV to facilitate efficient photoionization from the lower singlet state of the NV$^-$ (figure \ref{fig:AmbientSCC}). The SCC protocol presented in this work is largely inspired by Hopper et al. \cite{Hopper2016Near-infrared-assistedDiamond}, in their work, near-infrared (1064~nm) lasers were used with conventional optics to pump electron population in the NV into the singlet state and ionize the defect using the near-infrared laser. The main difficulty ionising from the NV lower singlet is cross-talk in the NV. The predicted energy gap from the lower singlet state to the diamond conduction band is about 570~nm (2.1 eV) \cite{Aslam2013Photo-inducedDetection}. In ambient conditions, a 570~nm laser has a high chance of exciting the NV triplet transition as well as the ionisation transition due to its broad absorption side-band \cite{Razinkovas2021PhotoionizationCalculations}. This cross-talk will create noise in the spin initialisation as well as the charge state readout. To avoid cross-talk, Hopper et al. used the 1064~nm laser to perform quadratic ionisation. Whilst the rate of triplet excitation is very low at this energy, the rate of ionisation is also low, reducing the chances of a successful spin-to-charge conversion. The optical spin contrast achieved was about 25\%, equivalent to that of conventional optical spin cycling techniques \cite{Jaskula2019ImprovedConversion, Balasubramanian2009}. 

The purpose of the electrode in our work is to apply an electric potential which shifts the NV energy levels relative to the diamond conduction band (figure \ref{fig:AmbientSCC}b)). Positive potentials will shift the energy levels closer to the conduction band, increasing photoionization probability whereas negative potentials will have the opposite effect. The change in the energy gap from the NV energy levels to the diamond conduction band is due to the different ways that the two structures are affected by the electrode potential. The diamond conduction band will feel an averaged effect of the potential which can be modelled using effective mass theory and the NV can will feel a much larger potential at a point in space as it is an atomic defect. This allows us to selectively change the ratio of photoionization to absorption cross-section in order to maximise or minimise ionization from any energy level in the NV. By adding a potential that shifts the NV energies towards the diamond conduction band, the singlet ionization energy gap can go below 2.0~eV. If the ionisation energy gap is small enough, then the triplet absorption cross-section will be effectively zero at the same energy, allowing for a higher probability ionization with low levels of cross-talk, resulting in a high optical spin contrast which we can calculate. 

In this paper, the experimental design is outlined and the SCC pulse sequence is described. The SCC contrast is calculated using rate equation modelling \cite{Fox2009QuantumIntroduction}, the results of which are then compared to another established ambient SCC protocol \cite{Jaskula2019ImprovedConversion}. The results are analysed and some alternate applications of the electrode are considered.  

\section{Experimental design and contrast}

We design our experimental setup with a cylindrical electrode fabricated over an NV with a thin wire connecting to a voltage source (figure \ref{fig:AmbientSCC}a)). The electrode and wire have a thin, insulating silicon oxide (SiOx) layer to prevent charges from moving from the diamond into the electrode. The electrode itself is a transparent indium-tin oxide (ITO) conductive layer that carries the electric potential and allows for optical illumination through the electrode. The electrode changes the energy levels in the diamond band structure which can be calculated using effective mass theory. The electrode will also alter the NV energy levels, but to a greater effect than the conduction band energy levels (figure \ref{fig:AmbientSCC}b)). More details of the electrode effects on the energy levels can be found in supplementary section \ref{App:1}  

The SCC protocol is shown in figure \ref{fig:AmbientSCC}c)/d), which is very similar to the protocol from Hopper et al. \cite{Hopper2016Near-infrared-assistedDiamond}. A green laser pulse initialises the NV into the $m_s = 0$ spin state and a combination of green laser and microwave excitation places the defect in the desired excited state spin level. Population in the $m_s = \pm1$ state preferentially decay via the intersystem crossing (ISC) into the NV singlet states, whilst $m_s = 0$ states preferentially decay to the ground state. When the electron population is in the relatively long lifetime lower singlet state ($\approx$200~ns \cite{Acosta2010OpticalDiamond}), a powerful ionisation pulse is applied to ionise the defect. In this setup, the $m_s = 0$ spin state is mapped to the NV$^-$ charge state and the $m_s = \pm1$ is mapped to the NV$^0$ charge state. At the end of the sequence, the charge state is readout with a 595~nm orange laser pulse. During the ionisation phase, the electrode has a positive potential to shift the energy levels closer to the diamond conduction band (figure \ref{fig:AmbientSCC}b)) and increase the ionisation rate.

If the electrode potential shifts the NV energy levels such that there is no cross talk available between the ionisation and triplet excitation energies, then the optical contrast can be calculated using rate equation modelling. The rate model consists of five NV energy levels (where the ${}^1$A$_1$ state is removed due to its low lifetime \cite{Ulbricht2018Excited-stateDiamond}) plus a single level for the ionised state. Transition rates between the states form a 6x6 matrix used in the rate model where the excitation/ionisation laser power and pulse times are free variables to optimise and the remaining constants can be found in literature \cite{Tetienne2012Magnetic-field-dependentImaging, Kalb2018DephasingNetworks}. Note that the microwave pulse that performs the spin manipulation is implicit in this model and is assumed to excite with 100\% probability \cite{Waldherr2011DarkNMR}. We also assume that the NV can be initialised into a particular spin state with 100\% fidelity. Whilst this isn't precisely true, it can be achieved with near-unity fidelity with careful manipulations \cite{Hopper2020Real-TimeReadout}. A more detailed description of the rate equation model can be found in supplementary section \ref{App:2}. 

In order to solve the rate equation we need the ratio of the triplet absorption to photoionization cross-section, $\sigma$ (figure \ref{fig:AmbientSCC}c)). This value tells us the probability of an electron being raised to the excited triplet state or ionised into the diamond conduction band during green excitation. Ideally, $\sigma$ should be as low as possible to prevent triplet ionisation of the $m_s = 0$ state, lowering spin contrast. This value is typically set by the intrinsic properties of the NV but can be theoretically altered with a two-step electrode potential, where the electrode is off during the green laser excitation phase but has a positive potential during the ionization phase, increasing the overall chance of ionisation (figure \ref{fig:AmbientSCC}d)). This is discussed in more detail in section \ref{disc}.   

To work out $\sigma$, the ratio of the absorption triplet cross-section and the triplet photoionization cross-section is taken for a given excitation energy. To achieve this we need to calculate the triplet absorption sideband of the NV and compare it to the photoionization cross section. Note that the ratio is the key data, not the cross-section itself, this means that the units are not important as long as they are the same across the two data sets. We studied the absorption cross section at room temperature by applying a similar calculation from Davies et al. \cite{Davies1974VibronicDiamond, Doherty2013TheDiamond, Razinkovas2021VibrationalDiamond} which uses the Frank-Condon theory of electronic and vibrational interactions during an electronic transition along with the Huang-Rhys model of transitions in a defect. The theory states that with a temperature-dependent electron-phonon coupling, the function that describes the vibrational overlap is given by: 

\begin{equation}\label{eqn:vib1m}
    F(\omega, T) = e^{-S} \sum_{i=1}^{\infty} \frac{S^i}{i!} F_i(\omega,T),
\end{equation}

where $S$ is the average Huang-Rhys factor which is a measure of the interaction of defect electrons with phonons in a crystal lattice \cite{HUANG1950TheoryF-centres, Kehayias2013InfraredDiamond} and $F_i(\omega,T)$ is the temperature-dependent function describing the vibrational overlap of an electronic transition with $i$ phonons. Equation \ref{eqn:vib1m} can be solved for any number of phonon interactions and different temperatures to create a phonon sideband. Details of this calculation can be found in supplementary section \ref{App:3}.  

\begin{figure}[!ht]
            \centering
            \includegraphics[width = 0.45\textwidth]{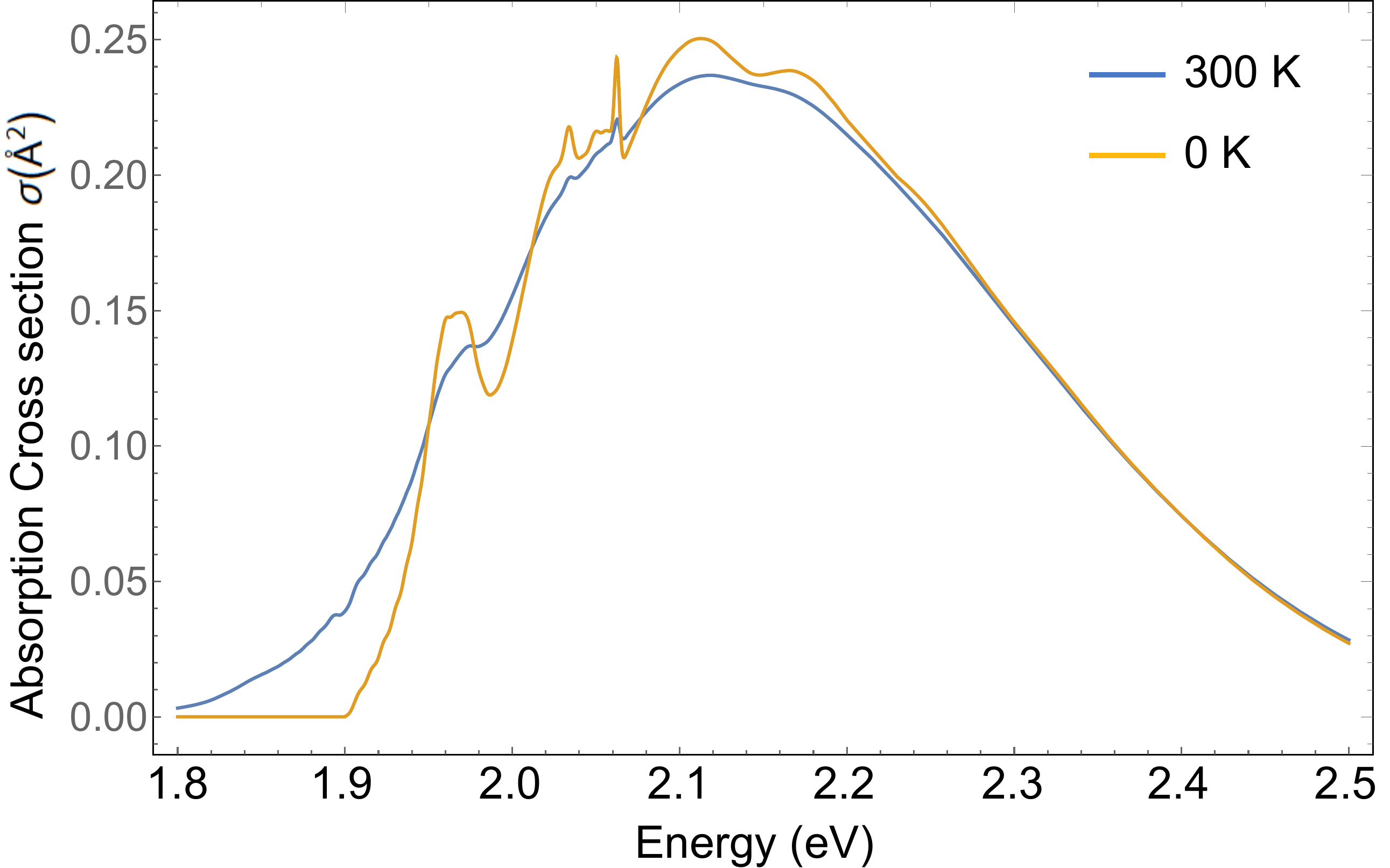}
            \caption{Plot of the absorption cross-section calculated using Huang-Rhys theory. The yellow curve is the solution at 0~K and is a match for the absorption cross-section data shown in Razinkovas et al. \cite{Razinkovas2021PhotoionizationCalculations}. The blue curve is the same calculation performed at 300~K. At higher temperatures, the data broadens, and lowers slightly in amplitude which is expected.}
            \label{fig:Absorption300K}
        \end{figure}

Figure \ref{fig:Absorption300K} shows the absorption spectrum for the NV across a range of photon energies with the zero phonon line (ZPL) omitted. The yellow curve shows the calculated absorption spectrum at 0~K which is the same as the data reported in Razinkovas et al. \cite{Razinkovas2021PhotoionizationCalculations}. The blue curve shows the same absorption spectrum calculated at 300~K. As expected, the data is largely the same, but the higher temperature electron-phonon interactions broaden the absorption spectrum. From this data, the cross-section for absorption can be compared to the cross-section for photoionization from Razinkovas et al. \cite{Razinkovas2021PhotoionizationCalculations} at 2.3~eV (532~nm) to obtain a ratio of $\approx$0.26. This ratio is then used as the $\sigma$ value in the rate equation modelling. 

\begin{figure}[!ht]
            \centering
            \includegraphics[width = 0.45\textwidth]{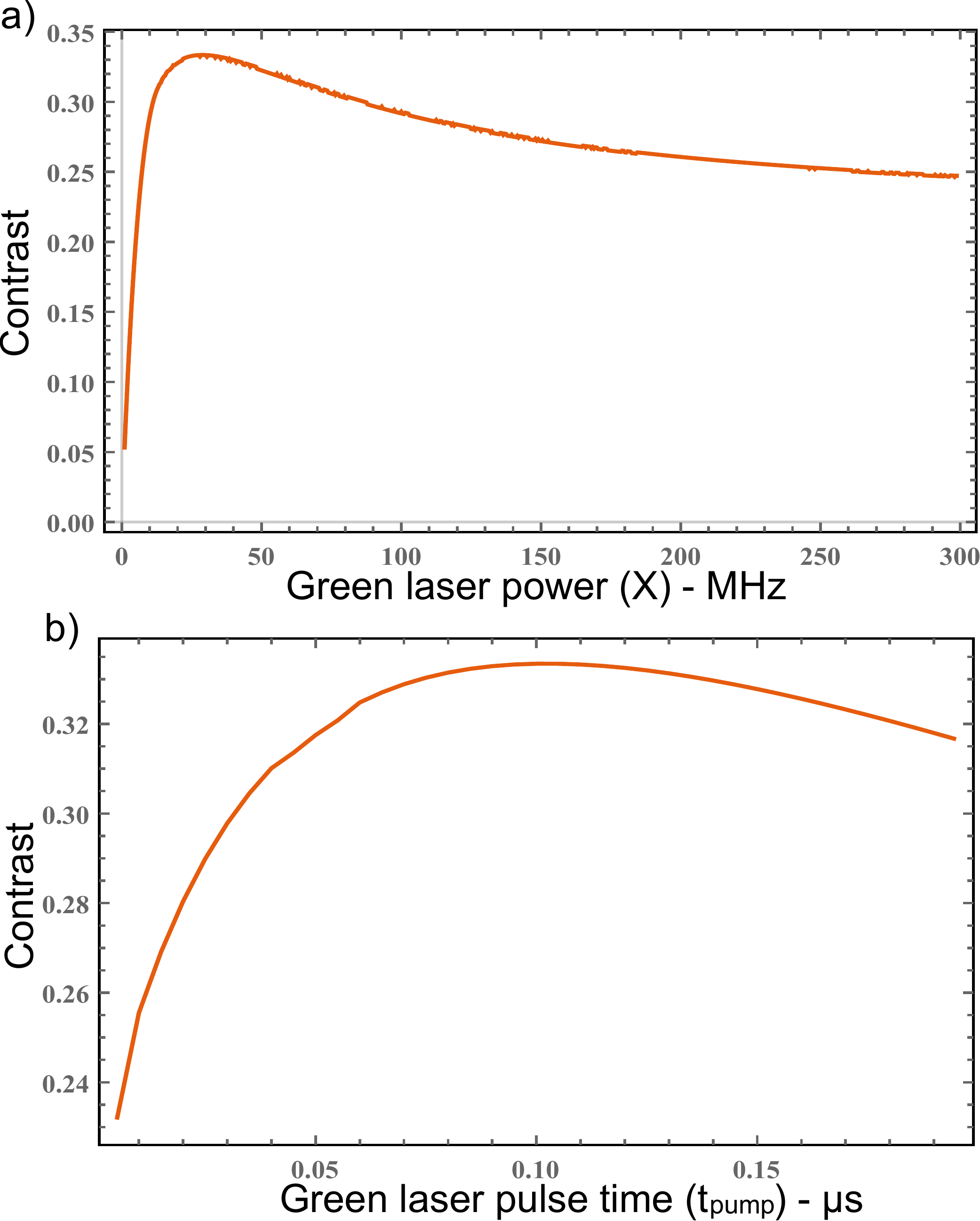}
            \caption{Plot of the optical spin contrast optimisation as a function of a) green laser excitation and b) green laser pulse duration. In both plots, the contrast steadily increases until reaching an optimum value of 33\% before decreasing.}
            \label{fig:conNoPulse}
        \end{figure}
        
Figure \ref{fig:conNoPulse} shows the optical spin contrast as a function of a) green laser power and b) green laser pulse duration, the remaining variables are optimised to give the largest contrast. The optimal green laser parameters maximises the shelving into the singlet state and the optimal ionisation laser parameters maximises the ionisation from the singlet state. The contrast is optimised at 33\% when the ionisation laser power and pulse time are maximised to fully ionise the defect and when the green laser has a medium laser power and pulse time to maximise shelving into the NV singlet states whilst minimising two photon ionisation. This result is higher than conventional NV optical cycling (25\% \cite{Balasubramanian2009}) but not necessarily higher than other established room temperature SCC protocols. 

\section{Comparison to established methods}

Another SCC protocol that operates at ambient conditions comes from Jaskula et al. \cite{Jaskula2019ImprovedConversion}. In their work, they ionise $m_s = 0$ population out of the ground state triplet state using two-step 637~nm quadratic ionisation. In this protocol, $m_s = \pm1$ population is shielded from ionisation via the intersystem crossing. Whilst their results are experimental, we can put their protocol into a similar rate equation and optimise to find a contrast that is directly comparable to our method, the details of which are in supplementary section \ref{App:4}. The optimised optical contrast we calculated from the Jaskula method was 37\%, which is a little higher than the experimentally achieved contrast reported at 36\% and higher again than the predicted contrast produced in our method. This is likely due to the higher ionisation cross section from the excited triplet compared to the singlet as the energy gap to the diamond conduction band is smaller. 

Given that the method proposed in Jaskula et al. does not require the added complexity of the electrode, it seems that their method is better in terms of both spin contrast and experimental simplicity. However, the technique proposed in this paper can be improved by using the same pulsing method proposed by Hopper et al. \cite{Hopper2016Near-infrared-assistedDiamond}. Contrast is improved by running the pulse sequence more than once and optimising the pulses for each run whilst leaving the spin manipulation and charge state readout processes unchanged. The idea is that during the pumping phase, some electron population in the $m_s = \pm1$ state will decay radiatively to the ground state instead of taking the ISC pathway to the singlet states. This permits a repetition of the pumping and ionisation phase to increase the probability of pumping population into the singlet for ionisation. In each pumping phase, the laser can have different parameters to optimise over to improve the process. Thus, if there isn't electron population in the singlet state in the first run, it might be in the second or third. In practice, as more sequences are added, the likelihood of ionising from the singlet increases which means that each subsequent run increases the contrast by smaller and smaller amounts as the probability of the electron population remaining in the triplet state gets smaller. The simulation showed a reasonable increase when adding up to three runs of the SCC protocol and that the increase in contrast for four or more runs is negligible. 

For experimental simplicity, the pulsing and ionization lasers only optimise over a new pulse duration for each run of the protocol and the laser powers are constant. Running this new protocol with three pulses gives a spin optical contrast of 42\%, this contrast is about 5\% higher than the spin contrast we calculated for the Jaskula et al. \cite{Jaskula2019ImprovedConversion}, however, requires more experimental apparatus to achieve (the electrode). It is worth noting that repeat pulsing is not possible using the method considered in Jaskula et al. as the ionisation and excitation occur out of the triplet states. 

To understand the optical contrast improvement, we can relate the change in spin optical contrast to sensitivity by using the DC magnetic sensitivity from Rhondin et al. as an example \cite{Rondin2014}.

\begin{equation}\label{eqn:sense}
    \eta_{dc} \sim \frac{1}{g \mu_B}\frac{1}{C \sqrt{n T_2^*}},
\end{equation}

where $g$ is the g-factor for the magnetic moment, $\mu_B$ is the Bohr magneton, $n = t_l*P$ is the optical collection efficiency defined by the total counts obtained from the NV, $P$, and the time of the readout $t_l$. $T_2^*$ is the NV electronic spin dephasing time and $C$ is the optical contrast. In principle, all the factors are constant except for the change in contrast, so the improvement in sensitivity is proportional to $1/C$. By subtracting the difference in the inverse contrast of one method to the other, we can predict that the Jaskula method offers an improvement in DC magnetic sensitivity of $\approx$1.2, compared to conventional optical cycling. The method in this paper offers an improvement in sensitivity of $\approx$1.6.

\section{Discussion} \label{disc}

\begin{figure}[!ht]
            \centering
            \includegraphics[width = 0.45\textwidth]{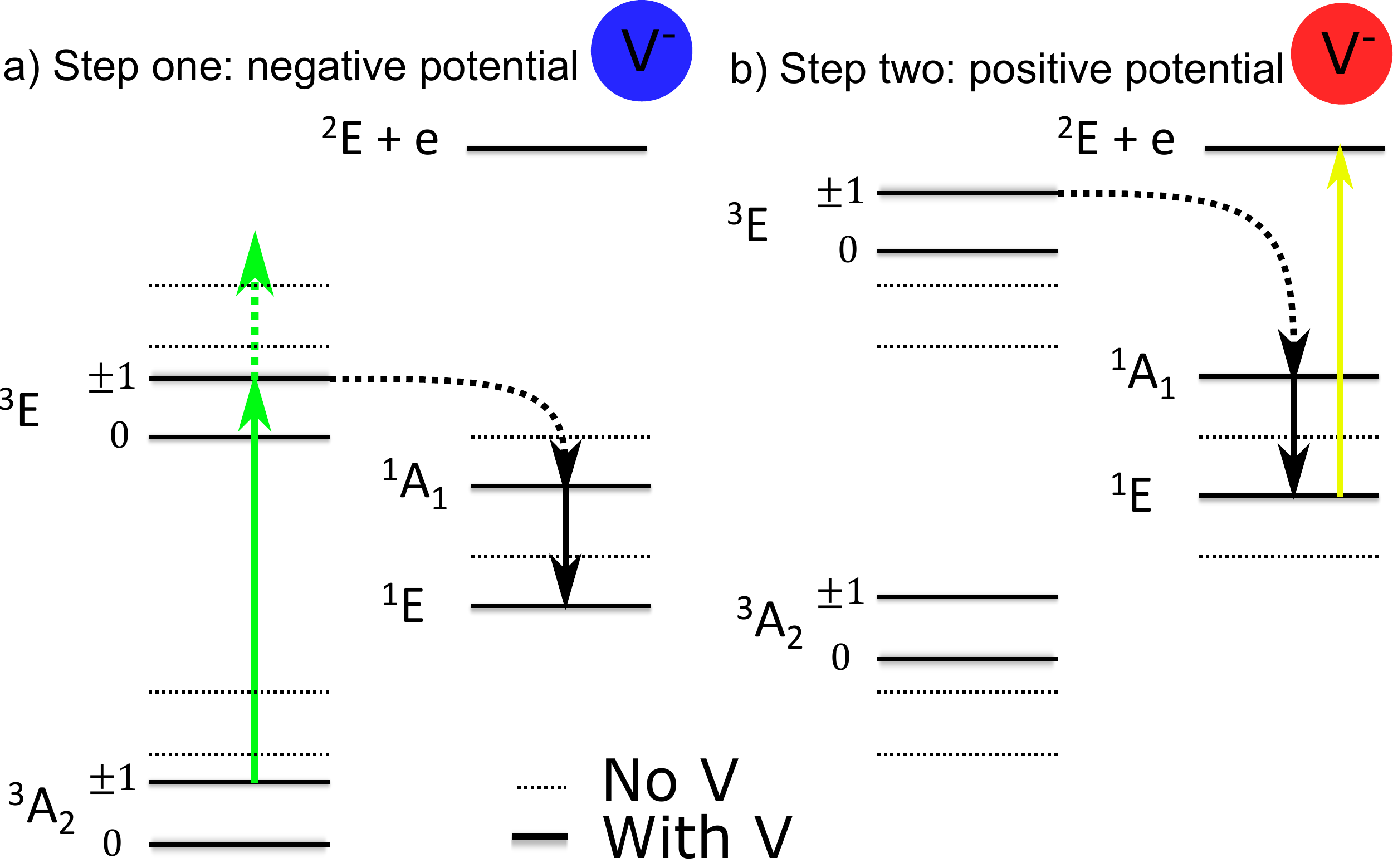}
            \caption{Images of the NV under different electrode potentials. In a), the electrode has a negative potential and shifts the energy levels away from the diamond conduction band (black lines compared to the dotted ones), reducing two photon ionisation due to the large energy gap from the excited triplet state. In b) the electrode has a positive potential, shifting the NV energy levels closer to the diamond conduction band, increasing the rate of quadratic ionisation as well as ionisation from the singlet states.}
            \label{fig:Epulse}
        \end{figure}

The electrode-based SCC method presented in this work, whilst showing a mild improvement over other protocols represents the highest optical spin contrast reported in the NV at ambient conditions. Whilst this method does modestly improve optical spin contrast, the real advantage lies in its potential capabilities. In the previous section, the electrode was only activated during the ionisation phase of the protocol to change the energy gap from the singlet to the ionised state. To alter the quadratic ionisation process from the triplet states, the SCC protocol would involve a two-step electrode potential as opposed to the single potential used in the previous section. An example of the two-step potential is illustrated in figure \ref{fig:Epulse}. In the two-step potential, the electrode would initially have a negative potential (blue) for the pumping phase (figure \ref{fig:Epulse}a)), shifting the NV energy levels away from the conduction band and reducing the probability of triplet ionisation. This is shown in figure \ref{fig:Epulse}a) by the green laser arrow which cannot make the gap from the ${}^3$E state to the ionised ${}^2$E + e state. In the second step of the process (figure \ref{fig:Epulse}b)), the ionisation phase, the electron is in the singlet state and the electrode would have its polarity reversed. This creates a positive potential (red), shifting the NV levels towards the conduction band and improving the rate of photoionisation from the singlet whilst reducing cross-talk. 

\begin{figure}[!ht]
            \centering
            \includegraphics[width = 0.45\textwidth]{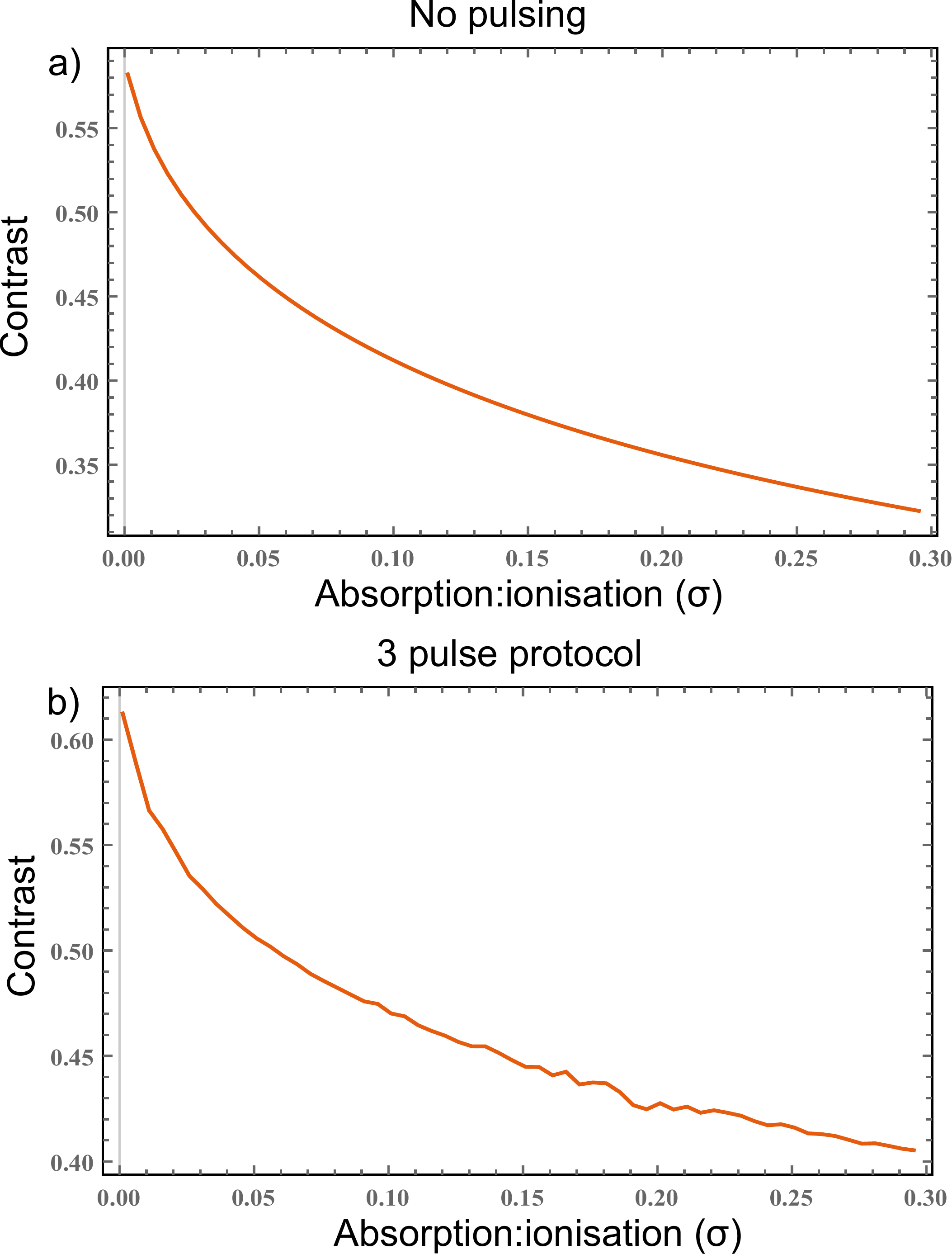}
            \caption{Plots of the optical spin contrast as a function of changing absorption to photoionization ratio $\sigma$. The left image a) is the optimisation process with no pulsing of the laser protocol, the right image b) is the same optimisation but with the pulsing sequence repeated and optimised three times. In both cases, the contrast is maximised when $\sigma$=0, i.e. no quadratic ionisation, with a steady decline in contrast with increasing $\sigma$.}
            \label{fig:sigma}
        \end{figure}

The effect of the negative electrode potential would be to change the ratio of triplet absorption to ionisation, $\sigma$. Figure \ref{fig:sigma} shows the optical spin contrast as a function of $\sigma$ optimised over the laser parameters. The contrast is maximised when $\sigma$ is zero, i.e. when there is no ionisation process out of the triplet states. When $\sigma$=0, the green laser pumping is maximised along with the ionisation laser and its pulse duration is short. In figure \ref{fig:sigma}a), there is no pulsing system and the contrast maximises at 58\%, in figure \ref{fig:sigma}b) there is a three pulse system and the contrast raises to 61\%. Without triplet ionisation, the main limitation of the optical spin contrast is the branching ratio at the ISC.

By using equation \ref{eqn:sense}, an optical spin contrast of 61\% translates to a 2.36-fold improvement in the NV sensitivity compared to conventional ODMR optical cycling techniques. However, using the electrode to alter the rate of quadratic ionisation has other added benefits. In a publication by Doi et al. \cite{Doi2016PureDiamond}, they postulate that improving the ratio of NV$^-$ to NV$^0$ during optical illumination would improve the number of photons the NV emits during a quantum operation. This was achieved in Doi et al. using dopants, but can be accomplished on single or ensemble NVs using an electrode by reducing the amount of triplet ionisation. Another consideration when using the electrode for charge state control is the effect it might have on the spin coherence time of the NV. In most modern theories of the NV, spin decoherence occurs as the spin state in the NV interacts with nearby paramagnetic defects in the diamond which cause a spin-flip in an NV electronic state \cite{Doherty2013TheDiamond, Herbschleb2019Ultra-longSpins}. Thus, most efforts to improve NV coherence time involve engineering diamond samples that remove these paramagnetic defects. However, if the NV ionises due to an unwanted ionisation from the spin triplet, then the spin information of the NV is lost along with its charge state, effectively creating decoherence. This can be compounded by the fact that an increase in ionisation and recombination attracts a local density of holes near the NV$^-$ which can interact and cause further ionizations even when there is no laser turned on for photoionization. Electrode-based charge state control can help reduce the number of photoionizations whilst providing a potential that repels local holes around the NV. 

The major issue with using the electrode to eliminate triplet ionisation is the unpredictable nature in which the NV energy levels react to such a large electrode potential. The idea is to shift the gap such that the minimum energy gap for quadratic ionisation is larger than the energy being used to excite the NV during the pumping phase. With a green 532~nm laser (about 2.3~eV), the electrode would have to shift the NV energy levels about 1.1~eV away from the conduction band. It is unclear what such a large potential would do to the NV energy levels or whether they would shift linearly with electric potential at such large values. This could be solved using density functional theory (DFT) calculations \cite{Nunes2001Berry-phaseInsulators}. 

\section{Conclusion}

The modelling performed in this paper shows a relatively straightforward way to improve NV optical spin contrast in ambient conditions through careful manipulation of optical pulses in an SCC protocol. With the pulsing mechanism in place, the 42\% contrast calculated promises a 1.6-fold increase in the NV DC magnetic field sensitivity which is a significant improvement on other mechanisms and has applications in quantum sensing, quantum computation and experiments to understand NV energy levels in a variety of conditions. The electrode itself has many other potential advantages such as improving optical photoluminescence and coherence time. As a result, there is a lot of future work to be done with the electrode. This includes the experimental realisation of the initial one step potential SCC protocol and experimental investigations of the possible two-step potential SCC protocol. Further theoretical modelling of the effects of high potentials on the NV energy levels is required as well as further experimental and theoretical investigations of the effects of the electrode on the NV charge state stability and its effect on the NV fluorescence and coherence. The results shown so far demonstrate concrete theoretical evidence of NV performance enhancement that is applicable in a variety of quantum technologies and the future work has great promise for a variety of alternative means of NV performance enhancement.  

\appendix
\section{Appendixes}\label{App}

\subsection{Shifting energy levels with the electrode}\label{App:1}

\begin{figure}[!ht]
            \centering
            \includegraphics[width = 0.45\textwidth]{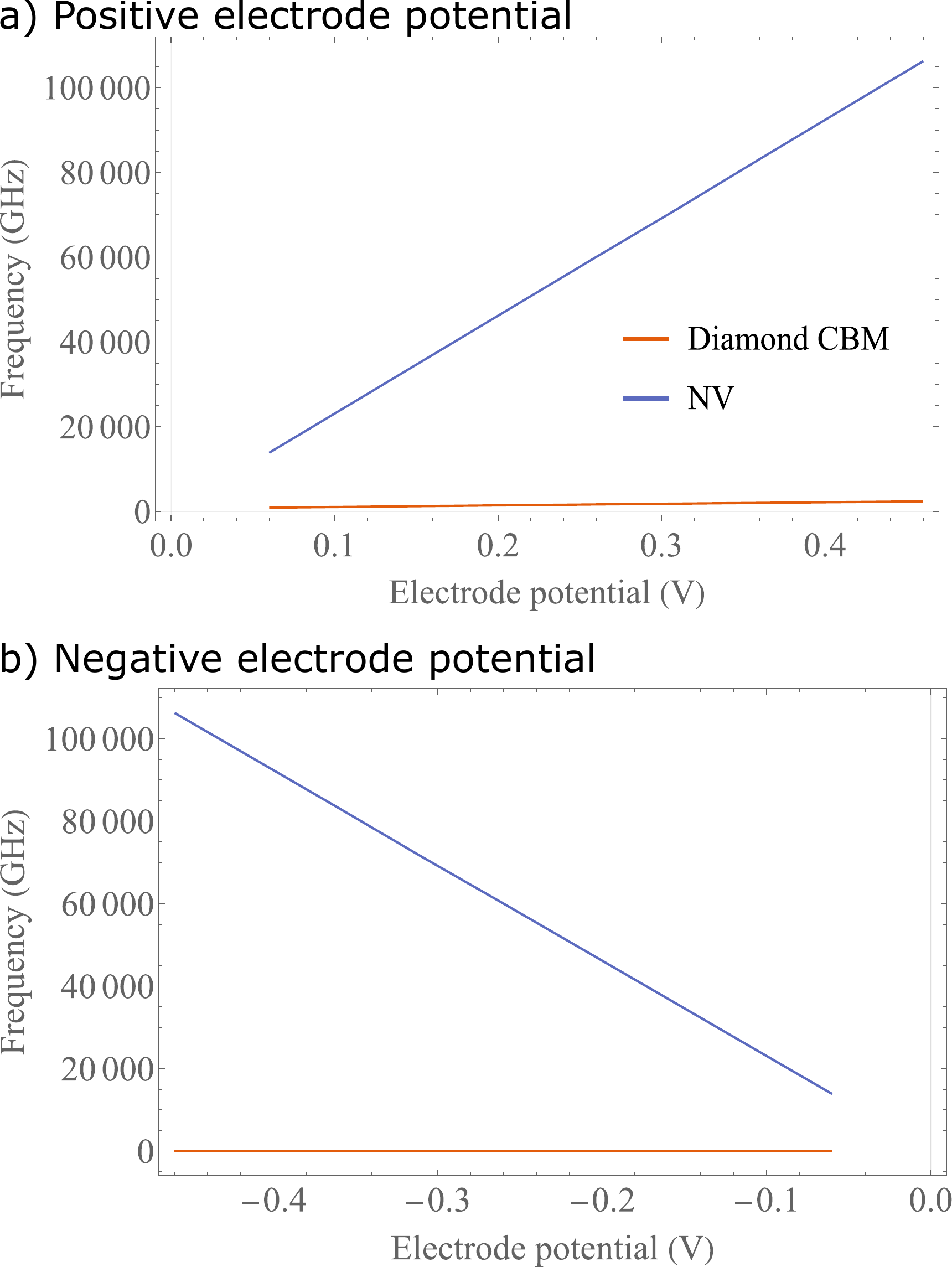}
            \caption{Plot showing the effect of the electrode on the electronic energy levels. The orange line is the shift of the NV energy levels due to the electrode when the NV is 10~nm below the diamond surface. The blue line is the shift in the diamond conduction band minimum state in the potential well which is calculated using effective mass theory. Figure a) is the effect for a positive electrode potential and figure b) is for the negative electrode potential. The NV states will shift more than the diamond states as the NV is a localised defect.}
            \label{fig:PotentialSplitting}
        \end{figure}

The electrode design is largely based off of another electrode SCC protocol \cite{Hanlon2021Spin-to-ChargeDiamond}, where an electrode was used to discretise the diamond conduction band at crygenic temepratures for resonant ionisations. At room-temperature, resonant ionisation is impossible due to electron-phonon broadening, however the potential will shift the diamond conduction band states using the same effective mass theory. We begin by solving the Schrodinger equation in the absence of an external potential: 

\begin{equation}\label{eqn:schrodinger2}
    \Big(\frac{-\hbar^2}{2} \Vec{\nabla} \cdot \overleftrightarrow{\frac{1}{m}} \cdot \Vec{\nabla} + V_c(\vec{k})\Big) \ket{F_n(\vec{r})} = E_n^c\ket{F_n(\vec{r})},
\end{equation}

where $E_n^c$ is the eigenenergy of the crystal system for a given energy level $n$, $V_c(\vec{k})$ is the crystal potential, $m$ is the effective mass tensor of the electron and $F_n$ is an envelope function which is related to the electron wavefunction by the following:

\begin{equation}\label{eqn:envelope2}
   \psi_n = F_n(\vec{r}) u_{\vec{k}}(\vec{r}) e^{i \vec{k} \cdot \vec{r}},
\end{equation}

where $u_{\vec{k}}(\vec{r})$ is the Bloch function for the state of the conduction band minimum in the bulk diamond unit cell and $\vec{k}$ is its associated wavevector. The exponent describes the phase difference when going between unit cells and we are considering $n$ conduction band minima as we expect new minima in different vector directions. Expanding out equation \ref{eqn:schrodinger2} to include the Bloch function gives: 

\begin{equation}\label{eqn:schrodinger3}
    \Big(T(\vec{r}) + V_c(\vec{k})\Big)F_n(\vec{r}) \mu_{\vec{k}}(\vec{r}) e^{i \vec{k}_{n} \cdot \vec{r}} = E_n^c F_n(\vec{r}) \mu_{\vec{k}}(\vec{r}) e^{i \vec{k}_{n} \cdot \vec{r}},
\end{equation}

where $T(\vec{r}) = \frac{-\hbar^2}{2} \Vec{\nabla} \cdot \overleftrightarrow{\frac{1}{m}} \cdot \Vec{\nabla}$, is the kinetic energy for a free electron with an effective mass $m$. For simplicity the Bloch function can be simplified to: $\mu_{\vec{k}}(\vec{r}) e^{i \vec{k}_{n}\cdot \vec{r}} = \ket{\phi(\vec{r})}$. Expanding equation \ref{eqn:schrodinger3} as a product rule whilst multiplying both sides of the equation by the complex conjugate $\bra{\phi(\vec{r})}$ gives the following: 

\begin{equation}\label{eqn:schrodinger4}
    \begin{split}
    \braket{\phi(\vec{r})}T(\vec{r})F_n(\vec{r}) + F_n(\vec{r})\bra{\phi}T(\vec{r})\ket{\phi(\vec{r})} + \\F_n(\vec{r})\bra{\phi}V_c(\vec{k})\ket{\phi(\vec{r})} = E_n^c F_n(\vec{r})\braket{\phi},
    \end{split}
\end{equation}

note that the envelope function is not acting on the eigenstates. The envelope function varies on distances much larger than Bloch function between diamond unit cells. Thus when considering small length scales of the diamond unit cell, one approximation being made is that $F_n$ is effectively constant and can be moved out of the inner product. Taking the inner product: 
 
\begin{equation}\label{eqn:schrodinger5}
    T(\vec{r})F_n(\vec{r}) + E_b F(\vec{r}) = E_n^c F_n(\vec{r}),
\end{equation} 

where $E_b$ is the energy of the Bloch function for the conduction band minimum. Another other key approximation being made is that an electrode confining potential isn't strong enough or varying enough on the scale of the unit cell such that the Bloch function depends on the potential. This means that the electrode potential can be added in a new Schrodinger equation as the Bloch function energy is independent of the electrode potential. Adding in the electrode potential gives the following:

\begin{equation}\label{eqn:schrodinger6}
    \big(T(\vec{r})+V(\vec{r})\big)F_n(\vec{r}) = E_n F_n(\vec{r}),
\end{equation}

where $E_n = E_n^c - E_b$ so that $E_n$ becomes the energy of the total wavefunction relative to the Bloch function. Equation \ref{eqn:schrodinger6} can be used to calculate the energy levels created by a confining electrode potential (ignoring broadening), By adjusting the external potential, the change in these levels can also be calculated. In contrast the NV shift in energy can be approximated as $eV_{NV}$ where $e$ is the electric charge constant and $V_{NV}$ is the potential at the point in the diamond where the NV is. Equation \ref{eqn:schrodinger6} describes the effect of the potential on a spatially averaged region of the bulk diamond whereas the effect of the potential on a single charged defect site is much more dramatic. For this reason, when the potential is applied to the electrode, the shift in the energy levels is expected to be much larger for the NV states than the diamond conduction band states which changes the difference in energy between the two states (figure \ref{fig:PotentialSplitting}).

For a positive electrode potential, the low lying conduction band states are confined within the potential well, creating discrete energy levels which increase linearly with the potential applied (figure \ref{fig:PotentialSplitting}a)). With a negative electrode potential, the conduction band states are not confined, so the effect of the potential to the diamond conduction band is averaged across the whole diamond. This means that the effect of the potential to the conduction band is negligible whereas the effect of the potential to the NV defect is the same as for the positive potential but with a negative slope (figure \ref{fig:PotentialSplitting}b)).  

\subsection{Rate equation modelling}\label{App:2}

\begin{figure}[!ht]
            \centering
            \includegraphics[width = 0.45\textwidth]{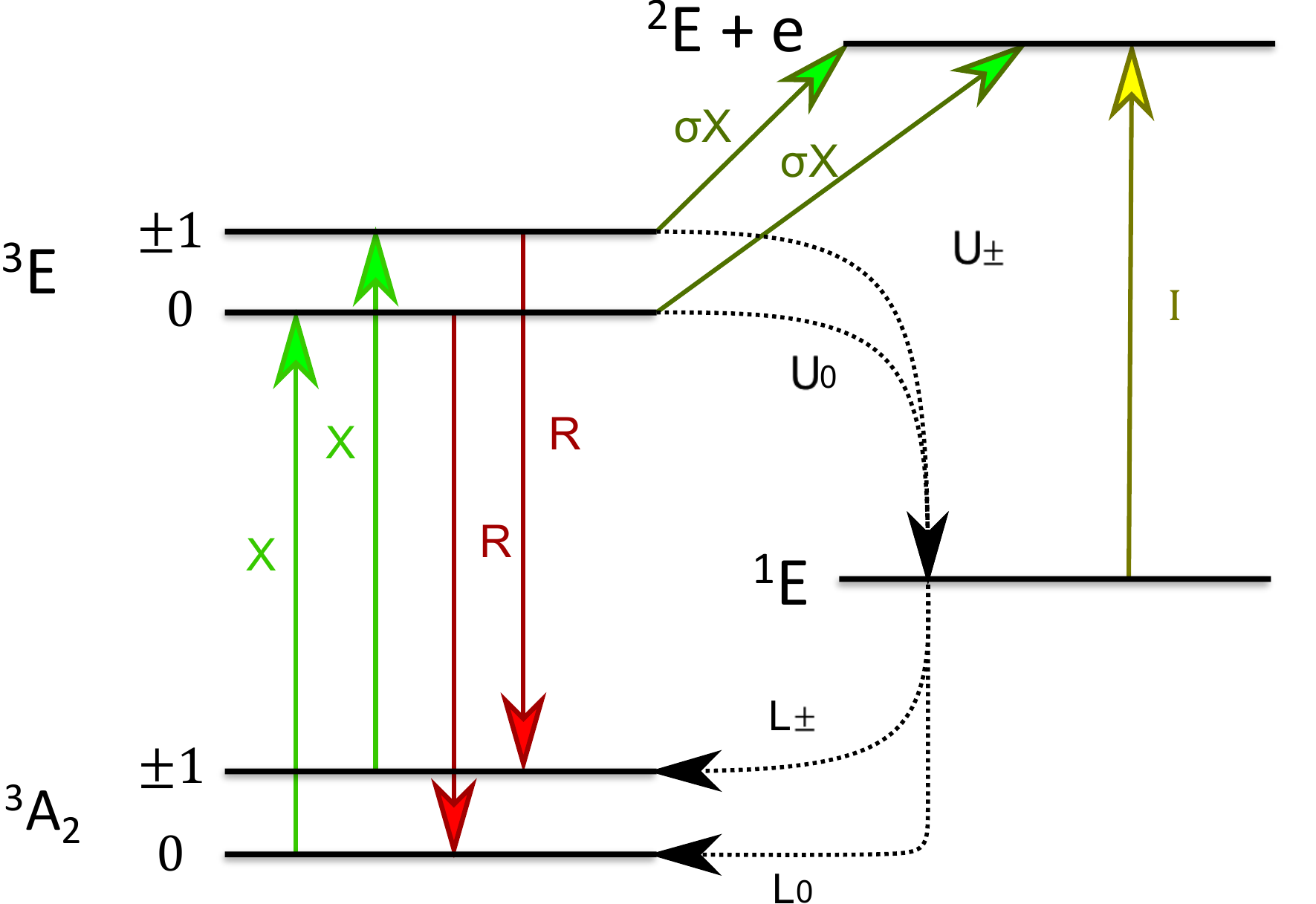}
            \caption{Image of the NV states considered in the rate equation model with the various transitions labelled including the ionised NV$^0$ $^2$E state with an electron in the conduction band. Note in the rate equation the excited singlet $^1$A$_1$ level is removed for computational simplicity. The solid arrows indicate optical transitions for both excitations and emissions whereas the dotted lines indicate non-radiative ISCs.}
            \label{fig:AmbientSCCrates}
        \end{figure}
        
Figure \ref{fig:AmbientSCCrates} shows the energy level diagram for the rate equation model being used in this simulation. This system features labels that represent the various excitation and relaxation pathways expected in the SCC protocol. The green $X$ terms represent the spin selective excitation from a green laser which excites the triplet manifold but also excites from the $^3$A$_2$ to the $^2$E$+e$ state via quadratic ionization. The ratio of absorption in the triplet to the quadratic ionization is given by $\sigma$:

\begin{equation}
    \sigma = \sigma_I / \sigma_A,
\end{equation}

where $\sigma_A$ is the absorption cross-section and $\sigma_I$ is the photoionization cross-section. Note that the microwave pulse that performs the spin manipulation is implicit in this model and is assumed to excite with 100\% probability \cite{Waldherr2011DarkNMR}. Also, note that the excitations are a single constant representing both the photon capture Einstein coefficient as well as the energy spectral density. The red $R$ terms represent the radiative emission pathway and the $U/L$ terms represent the spin-dependent non-radiative inter-system crossing (ISC). Note that in this model the excited $^1$A$_1$ level is removed for computational simplicity, as the decay rate is so fast from the $^1$A$_1$ to the $^1$E state compared to all other transitions that the pathway to the $^1$E state is effectively instantaneous \cite{Ulbricht2018Excited-stateDiamond}. Finally, the yellow $I$ terms represent the ionization pulse, which although is $\approx$570~nm without the electrode, should change in the presence of the electrode.

Each energy level has different sources/sinks that either populate or depopulate electrons from that level. These levels can be placed into a 6 entry vector for all the energy states and a 6x6 matrix that governs all the transitions between the states: 

\begin{widetext}
\begin{equation*}\label{eqn:SCCRate}
    \frac{d\vec{P}}{dt} = 
    \begin{pmatrix}
    -X & 0 & R & 0 & L_0 & 0\\
    0 & -X & 0 & R & L_{\pm} & 0\\
    X & 0 & -(R+U_0+\sigma X) & 0 & 0 & 0\\
    0 & X & 0 & -(R+U_{\pm}+\sigma X) & 0 & 0\\
    0 & 0 & U_0 & U_{\pm} & -(L_0+L_{\pm}+I) & 0\\
    0 & 0 & \sigma X & \sigma X & I & 0
    \end{pmatrix}\vec{P},
\end{equation*}
\end{widetext}

if we call the matrix in equation \ref{eqn:SCCRate} $\mathbf{M}$ and over some time period, $\mathbf{M}$ becomes constant then $\vec{P}$ can be solved using the following matrix exponential: 

\begin{equation}\label{eqn:SCCRateSolve}
    \vec{P}(t) = e^{\mathbf{M}t} \cdot \vec{P}(0),
\end{equation}

which can be achieved numerically. The initial condition $\vec{P}(0)$ is the vector state representing where the electron population is at the beginning of the protocol. If the protocol features multiple laser pulses with different pulse times then the sequence is expressed in the following way: 

\begin{equation}\label{eqn:SCCRateSolve2}
    \vec{P}(t) = e^{\mathbf{M}_{ion}t_{ion}} \cdot e^{\mathbf{M}_{pump}t_{pump}} \cdot \vec{P}(0),
\end{equation}

where $t_{ion}$ and $t_{pump}$ represent the times for the laser in the ionisation and pumping phases respectively whereas ${M}_{ion}$ and ${M}_{pump}$ represent the matrices during the ionisation and pump phases respectively.  

In this work, we assume that the electron in the NV can be initialised into a particular spin state with 100\% fidelity. Whilst this isn't precisely true, it can be achieved with near-unity fidelity with careful manipulations \cite{Hopper2020Real-TimeReadout}. As a result, the initial states in this system are the following:

\begin{equation}
    \vec{P}(0)|_{ms=0} = \begin{pmatrix}
    1 \\ 0 \\ 0 \\ 0 \\ 0 \\ 0
    \end{pmatrix},
    \vec{P}(0)|_{ms=\pm1} =
    \begin{pmatrix}
    0 \\ 1 \\ 0 \\ 0 \\ 0 \\ 0
    \end{pmatrix},
\end{equation}

where $\vec{P}(0)|_{ms=0}$ is the initial state with all the electron population in the first $m_s = 0$ state and $\vec{P}(0)|_{ms=\pm1}$ is the initial state with all the electron population in the second $m_s = \pm1$ state. The solution to equation \ref{eqn:SCCRateSolve} will give a vector of state probabilities based on the values in the matrix as well as the initial conditions. In this vector, the last entry corresponds to the probability of electron population existing in the ionised state at the end of the process. Thus, the spin optical contrast is calculated to be: 

\begin{equation}\label{eqn:contrastFinal}
    C = P_6|_{ms=\pm1}(t) - P_6|_{ms=0}(t), 
\end{equation}

where $P_6$ is the sixth entry in the vector which is calculated after the process has been solved for the spin zero case and the spin $\pm1$ case. 

The relaxation (R, U and L) rates are intrinsic properties of the NV that can be found experimentally. The radiative relaxation rate used in this study is R = 65.3~MHz \cite{Tetienne2012Magnetic-field-dependentImaging}, the upper ISC rates are: U$_0$ = 6.7~MHz and U$_{\pm}$ = 53~MHz, and the lower ISCs are: L$_0$ = 2.38~MHz and L$_\pm$ = 0.35~MHz \cite{Kalb2018DephasingNetworks}. For the excitation and ionisation rates, X and I are variables expressed as a function of the laser power in MHz. The X, and I terms are the variables to optimise over along with their associated pulse times. 

\subsection{Ratio of absorption to photoionization}\label{App:3}

To work out $\sigma$, the ratio of the absorption cross-section and the photoionization cross-section is taken for a given excitation energy. To achieve this we need to calculate the absorption sideband of the NV. The idea is to re-create the NV absorption data from Razinkovas et al. \cite{Razinkovas2021PhotoionizationCalculations} for ambient conditions and take the ratio at the energy of the green laser excitation ($\approx$2.3~eV). As phonons broaden the transition energies, the expectation is that at 300~K, the absorption cross-section will look also to broaden, however, the photoionization cross-section is already smooth and therefore isn't expected to change with temperature. We studied the absorption cross at room temperature by applying a similar calculation from Davies et al. \cite{Davies1974VibronicDiamond, Doherty2013TheDiamond} which uses the Frank-Condon theory of electronic and vibrational interactions during an electronic transition along with the Huang-Rhys model of transitions in a defect. The theory states with a temperature-dependent electron-phonon coupling, the function that describes the vibrational overlap is given by: 

\begin{equation}\label{eqn:vib1}
    F(\omega, T) = e^{-S} \sum_{i=1}^{\infty} \frac{S^i}{i!} F_i(\omega,T),
\end{equation}

where $S$ is the average Huang-Rhys factor. The Huang-Rhys factor is a measure of the interaction of defect electrons with phonons in a crystal lattice \cite{HUANG1950TheoryF-centres} and can be expressed with the following: 

\begin{equation}\label{eqn:HR}
    S = \int_0^{\Omega} \Big( 2n(\omega,T)+1\Big) f(\omega) d\omega,
\end{equation}

where $n(\omega,T)=1/(e^{E/k_b T}-1)$ is the temperature-dependent Bose-Einstein distribution of phonons and $f(\omega)$ is the low temperature one-phonon sideband function. This function describes the single electronic transition that can either create or annihilate a single phonon. It can be approximated by deconvolving the 0~K absorption sideband given by Razinkovas et al. \cite{Razinkovas2021VibrationalDiamond}. The temperature-dependent function describing the vibrational overlap of an electronic transition with one phonon can then be expressed as: 

\begin{equation}\label{eqn:vib2}
    F_1(\omega,T) = 
    \begin{cases}
    \big(n(\omega,T)+1\big)f(\omega) & \omega >0 \\
    n(-\omega,T)f(-\omega) & \omega <0  
    \end{cases},
\end{equation}

where the negative frequency $-\omega$ allows for the annihilation of phonons and sets $\omega$ as the transition frequency relative to the zero phonon transition frequency. 

To obtain the function for a two phonon interaction equation \ref{eqn:vib2} is then convolved with itself: 

\begin{equation}\label{eqn:convolve}
    \begin{split}
    F_2[\omega, T] = F_{1}(\omega,T)* F_1(\omega, T)\\ = \int_{\infty}^{\infty} F_{1}(\omega-x,T) F_1(\omega, T) dx,    
    \end{split}
\end{equation}

then to obtain any arbitrary number of phonon interactions, equation \ref{eqn:vib2} is simply convolved with itself the number of times required to obtain $F_i(\omega,T)$. By using the experimentally obtained Huang-Rhys factor of 3.49 \cite{Kehayias2013InfraredDiamond}, setting the temperature to 0~K and convolving the solution eight times the absorption spectrum shown in figure \ref{fig:Absorption300K} of the main paper is produced which is equivalent to the data from Razinkovas et al. \cite{Razinkovas2021PhotoionizationCalculations}. By keeping the same parameters but setting the temperature to 300~K we can obtain a new absorption side-band for the electron-phonon interactions at high temperature.  

\subsection{Comparison to Jaskula et al.}\label{App:4}

\begin{figure}[!ht]
            \centering
            \includegraphics[width = 0.45\textwidth]{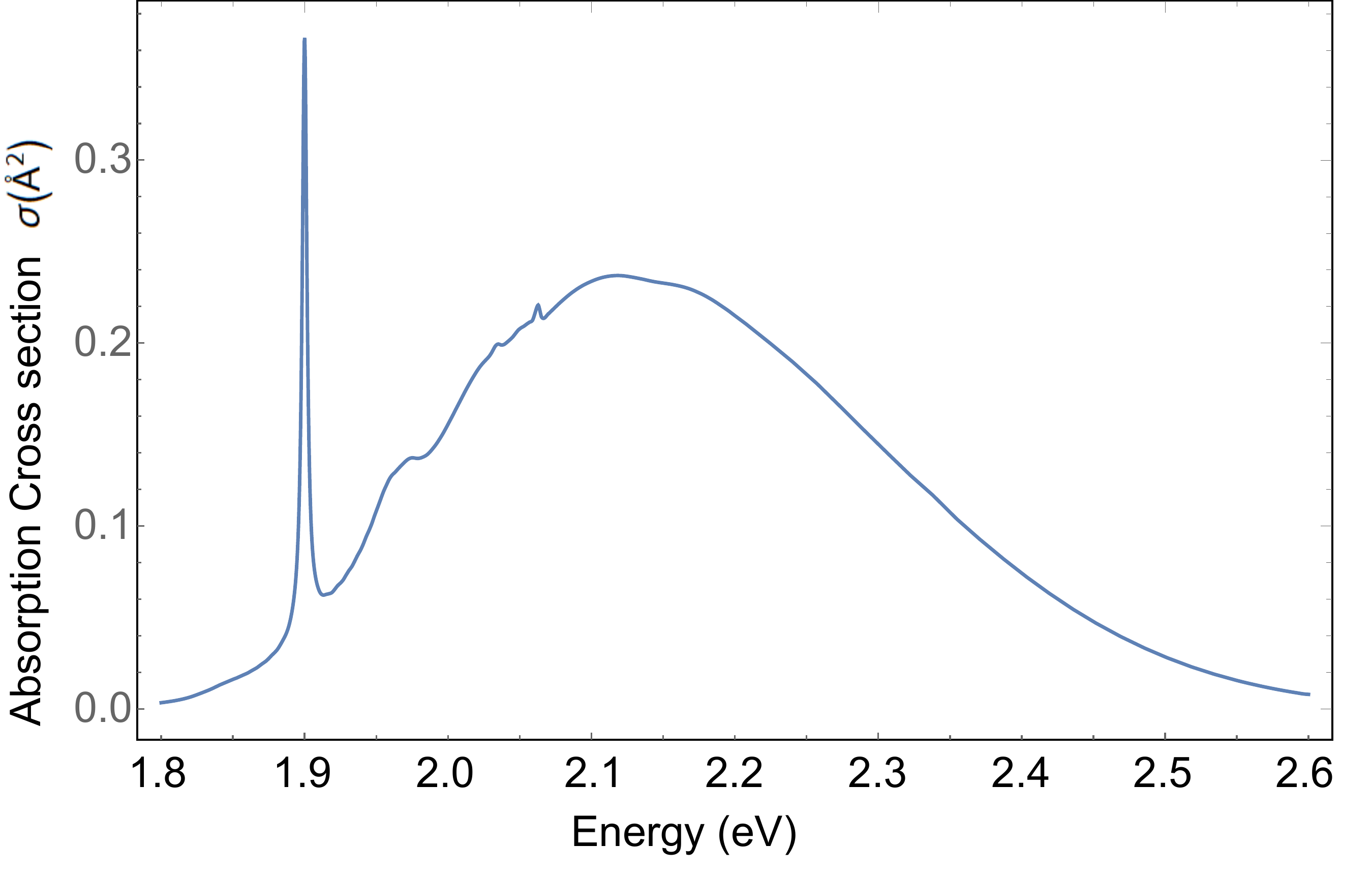}
            \caption{Plot of the same room temperature absorption cross-section used in figure \ref{fig:Absorption300K}, with the ZPL added in. The ZPL is added as a Lorentzian peak with a linewidth given by Fu et al \cite{Fu2009ObservationDiamond}. The peak is used to measure the absorption/ionisation ratio which is the new $\sigma$ value for the Jaskula et al. SCC contrast calculation.}
            \label{fig:AbsZPL}
        \end{figure}

To obtain the optical spin contrast of the method from Jaskula et al., rate equation modelling is used in a similar way as the previous section with some adjustments: 

\begin{equation}\label{eqn:SCCRateSolveJas}
    \vec{P}_{j}(t) = e^{\mathbf{M}_{ionJ}t_{ionJ}} \cdot e^{\mathbf{M}_{pumpJ}t_{pumpJ}} \cdot \vec{P}(0),
\end{equation}

where $\mathbf{M}_{ionJ}$ and $\mathbf{M}_{pumpJ}$ are the same matrices from equation \ref{eqn:SCCRate} but with I=0 and with different $\sigma$ values to reflect the change in the absorption to ionisation ratio for the different lasers. For $\mathbf{M}_{pumpJ}$, $\sigma$=0.16 which reflects the quadratic ionisation for the 594~nm laser designed to pump population into the singlet state. For $\mathbf{M}_{ionJ}$, the ratio needs to be found for the 637~nm laser used in ionisation which is the ZPL of the NV. Figure \ref{fig:AbsZPL} shows the same room temperature absorption spectrum from figure \ref{fig:Absorption300K} but with a ZPL added in as a Lorentzian whose linewidth at 300~K is 1~THz which is experimentally obtained from Fu et al. \cite{Fu2009ObservationDiamond}. From this figure and the photoionization cross-section Razinkovas et al. \cite{Razinkovas2021PhotoionizationCalculations}, the ratio was calculated to be 0.1.    

With the ratios for the different lasers found, equation \ref{eqn:SCCRateSolveJas} can be solved using the same parameters and optimising over the 594~nm pump laser as well as the 637~nm ionisation laser. 

\begin{acknowledgments}
The authors acknowledge the support from the Australian Research Council (DP 170103098). We would also like to thank Audrius Alkauskas for sharing absorption and photoionization cross-section data for our calculations and sharing his thoughts on the overall process. 

\end{acknowledgments}

\bibliography{1Main.bib}

\end{document}